%% file: BeverlySmith.tex
\documentclass[preprint2,longabstract]{aastex}

\def\chandra{\textsl{Chandra}}

\def\sun{\hbox{$\odot$}}
\def\farcm{\hbox{$.\mkern-4mu^\prime$}}
\def\farcs{\hbox{$.\!\!^{\prime\prime}$}}

\slugcomment{Accepted by the Astronomical Journal.}

\shorttitle{Chandra Imaging of NGC 7714/5}
\shortauthors{Smith et al.}

\begin{document}


\title{Chandra X-Ray Imaging \\ of the \\ 
Interacting Starburst Galaxy System NGC 7714/5:\\
Tidal ULXs, Emergent Wind, 
and
Resolved HII Regions }


\author{Beverly J. Smith}
\affil{Department of Physics, Astronomy, and Geology, East Tennessee
State University, Johnson City TN  37614}
\email{smithbj@etsu.edu}

\author{Curtis Struck}
\affil{Department of Physics and Astronomy, Iowa State University, Ames IA  50011}
\email{curt@iastate.edu}

\and

\author{Michael A. Nowak}
\affil{Center for Space Research, Massachusetts Institute of Technology, 77 Massachusetts
Avenue, Cambridge MA  02139}
\email{mnowak@alum.mit.edu}



\begin{abstract}
We present high spatial resolution X-ray imaging data for the
interacting galaxy pair NGC 7714/5 (Arp 284)
from the {\it Chandra} X-ray telescope.   
In addition to the unresolved starburst nucleus,
a variable point source with L$_X$ $\approx$ 10$^{40}$ erg~s$^{-1}$
was detected
1\farcs5 (270 pc) to the northwest of the nucleus, coincident 
with
a blue, extremely optically-luminous 
(M$_V$ $\approx$ $-$14.1) point
source on HST images.
Eleven other candidate point-like 
ultraluminous X-ray sources (ULXs) were also detected in
the vicinity of NGC 7714/5, two of which exceed 10$^{40}$ erg~s$^{-1}$.  
Ten of these appear to be associated
with interaction-induced features
but only two with star formation regions.
We also found
diffuse emission 
with L$_X$ $\approx$ 3 $\times$ 10$^{40}$~erg~s$^{-1}$
extending
11$''$ (1.9 kpc) to the north of the nucleus.
Its spectrum 
can be fit with either a 2-temperature
Mekal function (kT = 0.59 $\pm$ $^{0.05}_{0.06}$ keV and 
8 $\pm$ $^{10}_3$ keV),
or a 0.6 keV Mekal function plus a power law ($\Gamma$ = 1.8 $\pm$ 0.2).
The hard component may be due 
to high mass X-ray binaries (HMXBs) with
possible contributions from inverse Compton radiation,
while the soft component
is likely from a superwind.
Superbubble models imply an expansion age of $\approx$15 Myr,
supporting previous
assertions of
an intermediate-age nuclear
stellar population in addition to a 5 Myr starburst.
We also detected extended X-ray emission associated with four extra-nuclear
HII region complexes.
The emission from these HII regions and the nuclear starburst 
could be due to either an enhanced population of HMXBs
relative to Local Group galaxian averages, or 
to diffuse gas heated by winds from supernovae,
if the X-ray production efficiency L$_X$/L$_{mech}$
is high ($\approx$5$\%$).
To estimate L$_X$/L$_{mech}$, 
we collected published data for well-studied HII regions and superbubbles
in nearby galaxies.  For HII regions with ages $<$3.5 Myrs,
the median
L$_X$/L$_{mech}$ $\approx$ 0.02$\%$,
while for older star formation regions, 
L$_X$/L$_{mech}$ 
$\approx$ 0.2 $-$ 7$\%$.
Thus it is possible that 
gas heating by supernovae may be sufficient to account for
the observed X-rays from these HII regions.
In galaxies much more distant than NGC 7714, for example,
the Cartwheel galaxy,
HII region complexes
similar to those in NGC 7714 will be unresolved 
by \textsl{Chandra}, and will mimick
ULXs.  
No X-ray emission was detected from the Type Ib supernova SN 1999dn, with
an upper limit of $\approx$2 $\times$ 10$^{38}$ erg~s$^{-1}$.

\end{abstract}





\section{Introduction}

\subsection{X-Ray Emission from Starburst Galaxies}

The X-ray
properties of starburst galaxies tend to scale with star formation
activity 
\citep{fabbiano89,  read01,
ranalli03},
however, 
this X-ray emission is still not completely understood.
With the high spatial resolution of the \chandra\ telescope, we can 
now separate point sources from the diffuse
emission in starburst galaxies,
allowing these two 
components
to be studied with unprecedented detail.
\chandra\
observations have provided new information 
about the distribution,
temperature, and metallicity of hot X-ray-emitting
gas in starbursts, and in particular, about `superwinds':
high-velocity large-scale outflows from the nuclear
regions of starbursts
\citep[e.g.,][]{strickland00, 
strickland02,
strickland04a, strickland04b, 
martin02, 
rasmussen04}.
\chandra\ images have also revealed extended emission
associated with extra-nuclear star formation 
regions in both starburst and spiral galaxies
\citep{fabbiano03, zezas03, kuntz03, tyler04},
however,
this phenomenon is not well-quantified.

\chandra\ observations have also shown that star-forming galaxies
often contain numerous
`ultra-luminous' X-ray (ULX)
point sources 
(L$_X$ $\ge$ 10$^{39}$ L$_{\sun}$)
\citep{fabbiano01,
roberts02,
lira02, 
zezas02, zezas03, 
read03, smith03}.  The nature of these sources is still
unclear.
One suggestion is that they are 
``intermediate mass'' (100--$10^5$\,M$_{\sun}$) black
holes 
\citep{cm99, krolik04};
alternatively, they may be 
stellar mass black holes in X-ray binaries that are beamed 
\citep{king01, kording02}
or have super-Eddington accretion rates \citep{begelman02}.
More information about such sources,
particularly information about environment
and possible optical counterparts, is
needed to distinguish between these scenarios.

In order to study hot gas, superwinds,
and
ULXs in starburst galaxies, 
we require systems 
close enough that their X-ray emission can be resolved by
the 
\chandra\ telescope. 
Targeting starburst galaxies undergoing an interaction
with another galaxy can be especially useful if detailed
dynamical modeling of the interaction is available.
For
only a few starburst systems are sufficient data available for such a
comparison.  One such system is the peculiar galaxy pair NGC 7714/5
(Arp 284), which contains one of the best-studied examples of an
interaction-induced starburst.  
NGC 7714 is relatively
nearby at only 37\,Mpc (H$_0$ = 75 km s$^{-1}$ Mpc$^{-1}$; 1$''$ =
180\,pc).  

\subsection{The NGC 7714 Starburst and its Superwind}

Among nearby 
interacting systems not known to contain active nuclei, NGC 7714
has one of the most X-ray active
starbursts.  
Although the far-infrared, blue, and X-ray luminosities of
NGC 7714 are in the middle of the range for spiral galaxies,
its X-ray to blue
luminosity ratio L$_X$/L$_B$ is
almost an order of magnitude higher 
than typical values for spirals
\citep{david92}.
The starburst in NGC 7714 is often considered the prototype of
compact starburst nuclei (e.g., 
\citealp{weedman81}).
The UV, optical, and near-infrared spectrum of the
central $\approx$2$''$ of NGC 7714 has been the subject of
spectral synthesis studies \citep{gonzalez99,
lancon01},
which imply a
$\approx$5 Myr old starburst superposed on an older population.  

The unusual morphology of NGC 7714/5 provides strong constraints on
its dynamical history, making possible a detailed comparison of the
parameters of its starburst and the parameters of its interaction.
NGC 7714 has a partial ring, two
tails to the west and another to the east \citep[Figure 1;][]{arp66}.
The presence of both a partial ring and tails suggest an
off-center collision \citep{smith92}.
In 21 cm {\sc
Hi} maps \citep[Figure 2;][]{smith97}, a gaseous
counterpart to the western stellar tail makes a full loop
back to the bridge.  This
loop implies a prograde encounter relative to NGC 7714 
\citep{struck03}.
This bridge is rich in {\sc Hi} 
\citep
[M$_{\rm HI}$
$\approx$ 1.5 $\times$ $10^{9}~M_{\sun}$;][]{smith97}, 
with luminous H~II regions
\citep{bernlohr93, gonzalez95, smith97}.
The gas and young stars in the bridge are
offset to the north of an older stellar bridge.
Star formation is also seen in an arc northwest of the
nucleus, southeast of the nucleus, and in the inner western tail 
\citep{bernlohr93, gonzalez95, smith97}.
Molecular gas has been detected in the disk of NGC 7714,
but not elsewhere \citep{smith01}.

Using these constraints as well as kinematical information, we have
constructed a detailed hydrodynamical model of this system 
\citep{struck03}.
Our model predicts
significant mass transfer from NGC 7715 to NGC 7714, fueling the
central starburst and creating the inner southwestern tail.  The star
formation history of NGC 7714 predicted by our dynamical model is
consistent with that inferred from spectral synthesis models.
In the model, however, the predicted gas velocity dispersion in
the center of NGC 7714 is significantly less than in the real HI map,
implying contributions from a superwind. 
This is confirmed by 
ultraviolet spectroscopy with the Far-Ultraviolet Spectroscopic 
Explorer (FUSE), which shows a $\approx$900 km~s$^{-1}$ wind \citep{keel04}.

With the ROSAT X-ray satellite, 
\citet{papaderos98} found extended
X-ray emission near the center of NGC 7714, and concluded
that
this was due to hot
gas heated by a nuclear starburst.
Alternatively, the X-ray emission from the NGC 7714 nucleus may be due to 
X-ray binaries \citep{ward88}.
\citet{papaderos98} also detected a
second X-ray source 20$''$ east of the nucleus, which they
suggested was due to hot gas heated
by either infall of gas from the bridge or by a nuclear superwind.
More recent observations with the XMM-Newton X-ray
telescope by 
\citet{soria04} 
showed that this eastern source is variable and point-like, indicating 
that it is a possible ULX rather than hot gas.
\citet{soria04} also found that the XMM-Newton spectrum of the central region
had two components, a non-variable thermal plasma and a variable power law,
suggesting a possible obscured active galactic nucleus (AGN).
In contrast, Spitzer infrared spectroscopy shows no evidence
for an obscured AGN \citep{brandl04}.

\section{Observations and Data Reduction}

To study the X-ray emission from NGC 7714/5
at higher spatial resolution than previous observations, we 
observed 
NGC 7714/5 for 60 ksec on 24 January 2004 with the 
Advanced CCD Imaging Spectrometer
(\emph{ACIS})
backside-illuminated CCD chip S3 on the \chandra\ X-ray Telescope
\citep{weisskopf02}.  
This CCD has 1024 $\times$ 1024 0\farcs492 pixels.   To
minimize pile-up, we only read out 640 chip rows, giving an observed
field of view of 5\farcm2 $\times$ 8\farcm4.
This covers the entire NGC 7714/5 system, except for the end of
the eastern NGC 7715 tail.
The CCD temperature was $-$120$^{\circ}$, the frame time was
2.1 s, and the events were telemetered in FAINT mode.
In addition to the S3 chip (the focus of this paper), the S1, S2,
S4, I3, and I4 chips were also read out.

Initial data reductions were done using the {\bf CIAO} v3.1 
software\footnote{CIAO is the \chandra\ Interactive Analysis
of Observations data analysis system package, available
at http://cxc.harvard.edu/ciao.}.
We tested
for time intervals with high X-ray background,
and only included
events with ASCA grades 0, 2, 3, 4,
and 6, and removed standard bad pixels and columns.
The energy range was restricted to 0.3 $-$ 8 keV and then further 
divided into three ranges: 0.3 $-$ 1, 1 $-$ 2.5, and 2.5 $-$ 8 keV.
The binned images were adaptively smoothed using the {\bf CIAO} routine
{\it CSMOOTH}, using a minimal significance of signal-to-noise ratio (S/N)
= 2.5 and a maximal significance of S/N = 5.
The smoothed maps were then divided by a similarly smoothed
exposure map to convert into physical units and to remove
instrumental artifacts.  We experimented with exposure maps
created using different parameters, including both monochromatic
maps at various energies and weighted maps.  In all cases,
the basic morphology of the extended emission remained the same,
although the absolute flux level varied.  For our final 0.3 $-$ 8 keV
map, we used an exposure map weighted by the spectrum of the
27\farcs6 radius region centered on the NGC 7714 nucleus.
For the final 0.3 $-$ 1, 1$-$ 2.5, and 2.5 $-$ 8 keV maps, we used
monochromatic 0.8, 1.5, and 3 keV exposure maps, respectively.

\section{Results}

\subsection{Morphology}
In Figure 3, the final smoothed Chandra images in various energy ranges
are shown, with contours
of the broadband optical light from the
Digitized Sky Survey\footnote{The Digitized Sky Survey,
a compressed digital form of the Palomar Observatory Sky Atlas,
was produced at the Space Telescope Science Institute under US
Government grant NAGW-2166.  The National Geographic Society-Palomar
Observatory Sky Atlas (POSS-I) was made by the California Institute
of Technology with grants from the National Geographic Society.} 
(DSS) superimposed.
In Figure 4, the H$\alpha$ image from \citet{smith97}
is superimposed on the \chandra\ 0.3 $-$ 8 keV map.
An expanded view of the \chandra\ data, in contours, is superimposed on
H$\alpha$ in Figure 5.
The \chandra\ contours are overlaid on an archival Hubble Space
Telescope (HST) WFPC2 broadband red F606W image in Figure 6.
We registered the HST image by assuming the brightest \chandra\ source
is associated with the brightest source in the HST image; this also
provides good alignment for the extra-nuclear HII regions.

These figures show significant correlations between
the diffuse X-ray emission and star formation.
An extended arc of X-ray emission is detected along the inner
western tail of NGC 7714, coincident with H$\alpha$ emission (HII region
complex C; Figure 5).
Diffuse X-ray emission is also observed in the HII region complexes
northwest and north of the NGC 7714 nucleus
(Regions B and E; Figure 5).
Interestingly, the HII region complex between these two complexes
(Region D)
is not clearly detected in X-rays, even though it is brighter in H$\alpha$
than complex E to its east.
A fourth HII region complex $\approx$6$''$ to the southeast of the
nucleus (Region A) is visible in H$\alpha$ and X-rays, as well as
in the HST map (Figure 6), the radio continuum map of \citet{condon90}, and
the Br$\gamma$ image of \citet{kotilainen01}.
For HII regions A, B, C, and E, 
the radii inside of which 50$\%$ of the observed counts are contained
are 1\farcs5 (3 pixels = 270 pc), 2\farcs7 (5.2 pixels = 500 pc),
5$''$ (10 pixels = 900 pc), and 1$''$ (2 pixels = 180 pc), respectively,
compared to 0\farcs5 (1 pixel)
for the two brightest isolated point sources in the field,
ULX candidates \#4 and \#12 (see Section 3.3).

Figure 6 also shows diffuse X-ray emission extending 11$''$ (1.9 kpc)
to the northeast of the nucleus.  
Faint extended H$\alpha$ emission is seen in the inner
portion of this feature.
Comparison between the different images in Figure 3 shows that the
diffuse X-ray-emitting gas generally has a soft X-ray spectrum
compared to the point sources. 

\subsection{The Central Region of NGC 7714}

The high resolution \chandra\ map shows that there are two
central X-ray sources separated by 1\farcs5 (270 pc), with the brighter
source being to the southwest (see Figure 7a).
The 0.3 $-$ 8 keV X-ray luminosities for the nucleus and secondary
source are 
4.4 $\times$ 10$^{40}$ erg~s$^{-1}$ and
1.1 $\times$ 10$^{40}$ erg~s$^{-1}$, respectively (see Section 3.4).
Both sources are unresolved in the \chandra\ image, with the brighter
source being centered between two
image pixels.
In the registered HST F606W image,
there is a point-like optical counterpart within 0\farcs4 of
the fainter \chandra\ source (see Figure 7b).
This source is also visible in archival F300W,
F380W, F555W, and F814W HST images (see Section 4.3).

\subsection{Point Sources}

We used the {\bf CIAO} {\it WAVDETECT} routine to search for point
sources in the \chandra\ data.  Our observations are sensitive
to point sources at the distance of NGC 7714/5 with 0.5 $-$ 8 keV
luminosities (assuming isotropy) greater than approximately 
2 $\times$ 10$^{38}$ erg~s$^{-1}$, depending
upon spectral shape (flux density $^{>}_{\sim}$ 
10$^{-15}$~erg~s$^{-1}$cm$^{-2}$).
{\it WAVDETECT} detected 13 sources near NGC 7714/5 (see Figure 8 
and Table 1),
in addition to the nucleus of NGC 7714.

Sources \#3 and \#13 are relatively far from the optical
features of the galaxies and so are likely background.  
This leaves 11 candidate ULXs within the optical extent of
the galaxy.
Using source counts from the
Chandra Deep Field \citep{brandt01, rosati02} and extrapolating
to our bandpass as in \citet{humphrey03},
we expect $\approx$2 background sources above our flux 
limit within 
the 
optical extent of the galaxy.  Thus most of these 11 
point sources are likely ULXs associated with NGC 7714/5.
Two of these sources, \#4 and \#12, are very bright, with 
2657 and 1076 counts, respectively.
Source \#4, the object previously
studied by \citet{papaderos98} and \citet{soria04}, 
lies in the NGC 7714 ring, as does source \#5.
Sources \#8 and \#9
are in the inner western tail, while
source \#1 appears associated with the eastern NGC 7715 tail.
Sources \#10, \#11, and \#12
may be associated with the outer western tail.
Source \#2 appears to be in the disk of NGC 7715.
Sources \#6 and \#7 lie near the NGC 7714 bar.

Sources \#1, \#2, \#3, \#4, \#11, \#12, and \#13 and possibly
\#9 are marginally visible in the XMM-Newton image of
\citet{soria04}, while \#5, \#6, \#7, and \#8 are unresolved from
the nucleus.  Interestingly, source \#12, which is extremely bright
in the \chandra\ data, is faint in the XMM-Newton map, implying an
increase of a factor of $\approx$10 in luminosity since the XMM-Newton
observations.  It was not detected in the less sensitive
1994 ROSAT HRI observations
of \citet{papaderos98}.

Only two of the candidate ULX sources in Table 1
are associated with on-going star formation: \#8, which is associated
with HII region C, and \#9, which is coincident with
faint H$\alpha$ emission
on the \citet{smith97} H$\alpha$ map.
Source \#6 is close to but offset from star formation regions in the inner
disk of NGC 7714 (see Figures 5 and 6).
None of the other ULX candidates have optical counterparts on
archival HST or Digitized Sky Survey images.

No X-ray counterpart was found in the \chandra\ data for 
SN 1999dn,
a Type Ib (H~poor) supernova 14$''$ southeast of the nucleus 
\citep{deng2000}, or for the possible supernova 2$''$W, 5$''$N of the
nucleus claimed by \citet{mattila02}.
Our upper limit to the 0.3 $-$ 8 keV X-ray luminosity of these sources
is $\approx$2 $\times$ 10$^{38}$ erg~s$^{-1}$.

X-ray detections of optically-visible supernovae are uncommon.
The 
\citet{immler_lewin03}
compilation of X-ray supernovae contains only 15 detections,
with X-ray luminosities of 10$^{37}$ $-$ 10$^{40}$
erg~s$^{-1}$. 
Of these 15, none
were Type Ib and only three were Type Ic.
More recently, four additional Type Ic supernovae were detected,
bringing the count to seven \citep{pooley04b}.
The type Ib/c supernova 2001em in UGC 11794
was detected by Chandra with a very high luminosity
of $\approx$10$^{41}$ erg~s$^{-1}$
\citep{pooley04a}, significantly higher than our upper limit
for SN 1999dn.

\subsection{X-ray Spectra}

X-ray spectra for various regions in the NGC 7714/5 field
were extracted using {\bf CIAO} and spectral fitting was done using
the {\bf ISIS} data reduction package \citep{hd00}.
The best-fit parameters are given in Table 2, and the fits
are plotted against the observed spectra in Figure 9.
Before fitting the spectra, the data were rebinned into
20~counts~bin$^{-1}$.
The fits in Table 2 were made using the 0.5 $-$ 8 keV range;
similar results were found with 
0.3 $-$ 8 keV.
The quoted uncertainties are 90$\%$ confidence
level.
To estimate background counts, the deep ACIS observations of
blank fields provided by the ACIS calibration team were used
(file {\it acis7sD200-12-01bkgrndN0002.fits}), after reprojection
and scaling to our observations.
These background counts are consistent with, but higher S/N than,
counts from selected regions at the edge of our field.

The spectrum of the NGC 7714 nuclear region (within
a radius of 0\farcs86 = 150 pc) could be fit to a Mekal function
with kT = 9 $\pm$ $^{4}_2$ keV
and a column density n$_H$ of 1.5 $\pm$ $^{0.3}_{0.4}$ $\times$ 10$^{21}$ cm$^{-2}$.
The abundance is not well-constrained by this fit ($\le$2 solar).
An absorbed power law function also produces a good fit, with
photon index $\Gamma$ = 1.7 $\pm$ $^{0.1}_{0.0}$ and 
n$_H$ = 2.2 $\pm$ 0.2 $\times$ 10$^{21}$ cm$^{-2}$.
The spectrum of the secondary nuclear source (in a radius
of 0\farcs74) could be
fit to a power law with $\Gamma$ = 
1.9 $\pm$ $^{0.3}_{0.0}$.   This source requires
a column density similar to that of the nucleus, 2.0 $\pm$ $^{0.9}_{0.7}$
$\times$ 10$^{21}$ cm$^{-2}$.   
No Fe-K emission was detected, with an upper limit
of $^{<}_{\sim}$ 1.2 $\times$ 10$^{-14}$ erg~s$^{-1}$cm$^{-2}$.

We extracted a spectrum for the 
extended emission in the central region of NGC 7714
using a circular region with a 9\farcs5 radius centered
1\farcs5E, 2\farcs0N of the nucleus, excluding the regions
used for the nuclear and secondary source spectra and HII region A
6$''$ southeast of the nucleus.
The spectrum of this extended emission
is inconsistent with that of a power law or single temperature
gas.  It is well-fit by a two-temperature Mekal function,
with temperatures of 0.59 $\pm$ $^{0.05}_{0.06}$ keV and 8 $\pm$ $^{10}_3$ keV
and a column density of 5 $\pm$ $^4_3$ $\times$
10$^{20}$ cm$^{-2}$.
The spectrum can also be well-fit by a 0.6 keV Mekal function plus
a power law with $\Gamma$
= 1.8 $\pm$ 0.2.
The total 0.3 $-$ 8 keV luminosity of this extended emission is
3.5 $\times$ 10$^{40}$ erg~s$^{-1}$, with the
0.6 and 6 keV components
contributing 1.1 $\times$ 10$^{40}$ erg~s$^{-1}$ (33\%) 
and 2.3 $\times$ 10$^{40}$ erg~s$^{-1}$ (67\%), respectively.
For both the two thermal plasma fits and the thermal plasma plus power
law fits, we allowed the silicon abundance to be a free parameter
(constrained to be the same for both plasmas), to account for the
2\,keV feature seen in Figure~9.  
Both of these fits are consistent
with a Si abundance near solar ($1.2^{+0.6}_{-0.7}$ for the former,
and $0.9\pm0.7$ for the latter), and are marginally inconsistent with
the 0.25 solar abundance assumed for the other metals.
An enhancement of $\alpha$-elements such as Si
compared to other
elements is expected in starbursts,
since they are predominantly
produced by type II supernovae
 (e.g., \citealp{arnett95}).

We separated the extended emission into two zones, with the inner
being a 3$''$ radius region centered on the nucleus (excluding
the nucleus itself), and the outer
being the remainder
(excluding the secondary source and HII region complex A).
The spectra for both the inner and outer
zones requires
two components, and were well-fit with either
two thermal plasmas, or a thermal plasma and
a power law.
The temperatures
and spectral indices of these components were similar to those found
above.  The total $0.3 -8$\,keV luminosity for the inner region is $2.4
\times 10^{40}~{\rm erg~s^{-1}}$, with 17 $\pm$ $^{13}_7$\% 
coming from the soft
component and 83 $\pm$ $^1_{13}$\% coming 
from the hard component.   For the outer
region, the total X-ray luminosity is $1.2 \times 10^{40}~{\rm
erg~s^{-1}}$, with 66 $\pm$ $^{34}_{16}$\% from the cooler 
component, and 34 $\pm$ $^{16}_{34}$\% from the
hard component.  Thus the fraction of hard radiation decreases with
radius.
                                                                                
When separating the central diffuse spectrum into two regions, the
evidence for a greater than solar silicon abundance for the inner
region becomes stronger.  Again, we have constrained the Si abundance
to be the same in each component of the two plasma fits.  The 90\% lower
limits for the Si abundance are 1.6 and 2.9 solar for the two thermal plasma
and thermal plasma plus power law fits, respectively.  As also
discussed by Soria \& Motch (2004), there is some evidence for
abundance gradients when comparing the inner to outer region.  In
Figure 9, the silicon feature at 2\,keV is greatly diminished for the
outer diffuse region.  However, the error bars on the Si abundance for
the outer region are large and greatly overlap with those from the
inner region.  Thus, one can attribute the loss of the Si feature to
being predominantly caused by the loss of the hard X-ray component, rather
than specifically to an abundance gradient.

The extra-nuclear
HII region complexes had too few counts to fit X-ray spectra.
Assuming
a Mekal function with a temperature of 6.0 keV, N$_H$ = 2 $\times$ 10$^{21}$
cm$^{-2}$ (from the HI map of \citealp{smith97}),
and abundances
of 0.5 solar for HII region A and 0.25 solar for
the rest (from \citealp{gonzalez95}), we obtain
0.3 $-$ 8 keV luminosities for the HII region
complexes of 2 $\times$ 10$^{38}$
erg~s$^{-1}$ to 5 $\times$ 10$^{39}$ erg~s$^{-1}$ (see Table 3).
Lowering the assumed temperature to 0.65 keV decreases these luminosities by
a factor of $\approx$7, while   
lowering the column density by a factor of 10 decreases these values by 
$\approx$40$\%$.

Only ULX candidates \#4 and \#12 were bright enough to be
spectrally-fit. 
Source \#4, the bright source in the ring, could be well fit with
an absorbed power-law
with $\Gamma$ = 1.9 $\pm$ $^{0.1}_{0.0}$ and 
n$_H$ = 1.9 $\pm$ 0.3
$\times$ 10$^{21}$ cm$^{-2}$.
For source \#12, a single absorbed power law gave a reasonable
fit ($\chi$$^2_{\nu}$ = 1.24) for $\Gamma$ = 1.6 $\pm$ 0.2 and 
a low column density of 
n$_H$ = 6 $\pm$ 5 $\times$ 10$^{20}$ cm$^{-2}$.
Adding a thermal second component does not significantly improve this fit.
For the fainter point sources in Table 1,
we calculated their X-ray luminosities assuming $\Gamma$ = 1.8 and
n$_H$ = 1.8 
$\times$ 10$^{21}$ cm$^{-2}$, and assuming that they are at the distance
of NGC 7714.  

For the sources with too few counts to fit spectra, plotting
X-ray colors on a color-color diagram can provide some indication
of the photon index and column density.
Following the lead of \citet{prestwich03},
for the fainter candidate ULXs and the HII regions in NGC 7714/5, 
we 
plot the soft X-ray color (M - S)/(S + M + H) against the
hard X-ray color (H - M)/(S + M + H) in Figure 10, where S is the counts in
the 0.3 $-$ 1 keV band, M is the counts in the 1 $-$ 2 keV band,
and H is the counts in the 2 $-$ 8 keV band.
We have over-plotted curves indicating the colors of absorbed power
laws with column densities from 10$^{20}$ $-$ 10$^{24}$ cm$^{-2}$,
and photon indices $\Gamma$ = 1 $-$ 4.
Figure 10 shows that these objects have colors typical
of luminous point sources in other galaxies
\citep{prestwich03, schwarz04}.
The inferred column densities and photon indices for ULX candidates
\#4 and \#12 
are reasonably consistent with those obtained from 
the spectral fits (Table 2).
The HII regions have somewhat softer
X-ray colors than the point sources,
and regions B, C, and E appear somewhat less absorbed than
HII region A, as expected from their locations
in the galaxy.

\subsection{Variability}

We have assessed the variability of the two nuclear sources and the
remaining candidate ULX sources, \#1-13, with the `Bayesian Blocks'
algorithm of Scargle et al. (2004, in prep.)\footnote{The specific
implementation that we used was from the {\tt S-lang ISIS Timing
Analysis Routines}, {\tt v.0.4.2}, available and described at {\tt
http://space.mit.edu/CXC/analysis/SITAR}.}.  The algorithm takes an
event lightcurve and finds the `optimal partitioning' into
piecewise-constant `blocks' of uniform rate. Thus, a variable
lightcurve is represented as a series of step functions.  The
partitioning of the lightcurve is governed by a prior probability for
the count rate in a block (here we choose this probability as $\propto$
e$^{(-{\Lambda}/{\Lambda}_0)}$, where $\Lambda$ is the count rate and
$\Lambda_0$ is the mean rate over the entire lightcurve), and a prior
probability for the number of blocks, $N_b$, that is $\propto
\gamma^{N_b}$.  Thus, $1-\gamma$ is, very roughly, similar to a
`significance level' for \emph{each} block; however, it should not be
interpreted as a strict quantitative assessment of significance.  The
advantage of this algorithm is that it makes no prior assumption
concerning the time scales of variability -- either long or short time
scale blocks can be found -- nor on the location or number of the block
boundaries.  In the results described below, we find evidence for both
short and long time scale variability.

We first apply the variability search to the primary nucleus.
At
values of $1-\gamma < 0.96$, the lightcurve is divided into
three blocks.  The first block has a duration of approximately
40\,ksec and a count rate of $(2.72\pm0.09)\times10^{-2}$\,cps, while
the last block has a count rate of $(2.31\pm0.10)\times10^{-2}$\,cps
and a duration of approximately 20\,ksec.  The intervening block is
only 128\,sec wide and contains 13 photons, for a nominal count rate
of $0.10\pm0.03$\,cps.  Viewed individually, this nearly factor of
four flare is over 99.99\% significant; however, accounting for the
fact that there are 470 independent 128\,sec intervals in our data,
the significance level is 96\%, in good agreement with the simple
interpretation of $1-\gamma$.  The two large blocks, on the other
hand, are detected for values of $1-\gamma < 0.995$, in
reasonable agreement with the fact that their count rates are
approximately 4$\sigma$ apart from one another.

The secondary nucleus shows even more statistically significant
variability.
For values of $\gamma > 4\times10^{-7}$, the lightcurve
is divided into two blocks.  The first block is of duration 46\,ksec
and has a rate of $(8.7\pm0.4)\times10^{-3}$\,cps, while the second is
of duration 14\,ksec and has a rate of
$(1.5\pm0.1)\times10^{-2}$\,cps.  Substructure is only found for
$1-\gamma < 0.85$, and therefore is only marginally significant.  In
Figure~11 we show the lightcurves for the primary and secondary
nucleus, binned to 1\,ksec and 2\,ksec, respectively, and we overplot
the `two block' decompositions found in our variability searches.

We also have used the Bayesian Blocks algorithm to search all 13
candidate ULX sources for signatures of variability.  Only the second
brightest source, ULX \#12, shows any significant variability.  For
values of $1-\gamma < 0.92$, the ULX \#12 lightcurve is decomposed
into five blocks, with the first and last blocks having rates of
$(1.7\pm0.1)\times10^{-2}$ cps.  These blocks are separated by a
19.2 sec interval with six photons (divided into 3 blocks), which is
approximately 15 times the mean count rate. This represents a single
realization probability of occurence of $6\times10^{-7}$. Factoring in
that there are 3128 independent 19.2\,sec intervals in our data, this
yields a 99.6\% significance level for this source.  Even accounting
for the fact that we have searched for variability in all 13 ULX
lightcurves, this is still 95\% significant.  We further note that all
6 events come from a $2\times3$ pixel region located at the center of
the source position, and thus are indeed identified with the ULX
source.

\section{Discussion}

\subsection{The Nuclear Extinction}

The best fits to the X-ray spectra of the NGC 7714 nucleus and
the secondary source near the nucleus both
require reasonably high column densities of 
N$_H$ $\approx$
2.2 $\times$ 10$^{21}$ cm$^{-2}$.  This is higher than
the sum of the 
HI absorption column density of 5 $\times$ 10$^{20}$
cm$^{-2}$ \citep{smith97} plus the Galactic extinction of $\approx$ 3
$\times$ 10$^{20}$ cm$^{-2}$
\citep{burnstein82, stark92, schlegel98}.
Our X-ray-derived column densities, 
however,
are consistent with extinctions obtained
from optical and near-infrared hydrogen recombinations,
A$_V$ $\approx$ 1 $-$ 2 \citep{puxley94, calzetti96, kotilainen01},
which imply N$_H$ $\approx$ 2 $-$ 4 $\times$ 10$^{21}$ cm$^{-2}$
using the standard ratio of N$_H$/E(B $-$ V)
= 5.8 $\times$ 10$^{21}$ cm$^{-2}$/mag \citep{bohlin78}.
This suggests additional extinction from dust associated with
molecular gas in addition to that associated with
HI.
HI emission line maps show a central depression in NGC 7714
\citep{smith97}, while 2.6 mm CO interferometric
maps show a central $\approx$5$''$ $\times$ 10$''$ source 
elongated southeast-northwest 
\citep{ishizuki93}.
The CO surface brightness for this source is $\approx$25 K~km~s$^{-1}$,
which corresponds to 
N$_{H_2}$ = 6.5 $\times$ 10$^{21}$ cm$^{-2}$,  
using the standard Galactic N$_{H_2}$/I(CO) ratio of 
\citet{bloemen86}.   This is consistent with column densities
based on the X-ray and hydrogen recombination line data.

\subsection{The Origin of the X-Rays from the Nuclear Starburst}

Possible contributors to the X-ray emission from a nuclear starburst
include supernovae, O stars, Wolf Rayet (WR) stars, hot gas, high mass
X-ray binaries (HMXBs), ULXs, and an obscured AGN.
In the NGC 7714 nucleus, supernovae and hot stars are likely
to be minor factors.
In their stellar synthesis study of 
the inner 0\farcs9 (150 pc) radius of 
NGC 7714, 
\citet{gonzalez99} obtained a starburst age
of 5 Myr, with a supernova rate
of 0.007 yr$^{-1}$.
With new optical and near-infrared data, \citet{lancon01}
updated this model, and concluded that an older population
was also required.
One possibility is a second intermediate-age ($\approx$20 Myr) burst,
which
boosts the predicted supernova rate
by a factor of $\approx$10.
Assuming an average L$_X$ $\approx$ 2 $\times$ 10$^{36}$
erg s$^{-1}$ per supernova and a time scale of $\approx$2 $\times$
10$^{4}$ yr for supernova remnants \citep{cowie81, williams97},
the 
X-ray luminosity from supernovae is $\approx$3 $\times$
10$^{39}$ erg~s$^{-1}$, $<$10$\%$ of the
absorption-corrected
\chandra\ X-ray luminosity of the nucleus of
4 $\times$ 10$^{40}$ erg~s$^{-1}$.
Typical X-ray luminosities for
O stars and WR stars
are
$\approx$10$^{31 - 32}$ erg~s$^{-1}$
and $^<_-$10$^{33}$ $-$ 10$^{35}$ erg~s$^{-1}$, respectively
\citep{sciortino90,
roncel03}.
The NGC 7714 nucleus has $\approx$2000 WR stars and $\approx$16,600 O stars
\citep{gonzalez99}, implying a total L$_X$ $\approx$ 10$^{36}$ $-$ 10$^{38}$
erg~s$^{-1}$ from these stars, $<$1$\%$ of the observed X-ray luminosity.

The contribution from 
hot gas heated by injection of mechanical energy from
supernovae is more uncertain.
Assuming each supernova contributes 
10$^{51}$ erg of kinetic energy, the \citet{lancon01} supernova rate
implies
a mechanical
luminosity from the supernovae of
$L_{SN}$ $\approx$ 2 $\times$ 10$^{42}~{\rm
ergs~s^{-1}}$,
50$\times$ larger than the 
\chandra\ X-ray luminosity for this region.  This
indicates that mechanical energy from supernovae
is sufficient to account for the X-ray luminosity
if the efficiency of converting this mechanical
energy into X-rays 
L$_X$/L$_{mech}$ is 
greater than $\approx$2\%.
We discuss 
this X-ray production factor
L$_X$/L$_{mech}$ further in Section 4.5; at the present
time it is quite uncertain.

Another possible source of nuclear X-ray emission 
is HMXBs, which are associated with a young stellar population.
For Local Group galaxies,
\citet{helfand01} found ratios of the total X-ray luminosity
from HMXB, L$_X$(HMXB), to 
the 
number of O stars 
of
2 $-$ 20 $\times$ 
10$^{34}$ erg~s$^{-1}$/star.
These ratios 
imply that
$<$10$\%$ of the nuclear X-ray flux from NGC 7714 is due to HMXBs.
This limit
is uncertain, however, since the \citet{helfand01} ratios are averaged over
entire galaxies with a range of stellar ages.
In an single coeval star formation region, one would expect
no HMXBs until stars start evolving off the main sequence, after which
L$_X$(HMXB)/L$_{H\alpha}$ will increase.
The
evolution
of the HMXB luminosity function 
and 
L$_X$(HMXB)/L$_{H\alpha}$ 
in starbursts is still not well-determined, thus how much HMXBs 
contribute to the X-ray emission from the nuclear region is still uncertain.

Alternatively, the unresolved nuclear emission may be caused by either
a low-luminosity AGN or one or more ULX-like objects.
This idea is supported by
the possible variability of the nucleus, which suggests that the emission
is dominated by a single source or a few sources.
The X-ray spectrum and luminosity
is consistent with those of both AGNs and ULXs
\citep{schwarz04}, however, as noted earlier, Spitzer infrared
spectra show no evidence for an obscured AGN
\citep{brandl04}.  
The upper limit to the luminosity of the Fe-K line from
the nucleus is 2 $\times$ 10$^{39}$ erg~s$^{-1}$.  Comparing with
the far-infrared luminosity of 8.3 $\times$ 10$^{43}$ erg~s$^{-1}$
\citep{david92} gives L$_{Fe-K}$/L$_{FIR}$ $\le$ 2.4 $\times$ 10$^{-5}$,
1 $-$ 2 orders of magnitude lower than that of typical Type 2 Seyferts
\citep{ptak03},
which argues against an obscured AGN. 
Furthermore, 
the 128 second flare followed
by a diminution of the X-ray flux is similar to the X-ray flares seen
in the microquasar GRS 1915+105 immediately preceding radio ejection
events \citep[e.g.,][]{belloni97}.  
If this flare is real,
then assuming that its time scale must be longer
than a dynamical time scale, e.g., the orbital period at the innermost
stable circular orbit of a Schwarzschild black hole, limits the compact
object mass to
$^{<}_{\sim}$3 $\times$ 10$^5$ M$_{\sun}$.
This mass limit also argues
against an obscured AGN.

\subsection{The Secondary Nuclear Source}

The secondary nuclear source in NGC 7714 may be an ULX, having an
X-ray luminosity and power law
spectrum consistent with that of the ULXs studied by \citet{schwarz04}.
We obtained HST WFPC2 magnitudes for the possible 
optical counterpart
to this source using the IRAF
software\footnote{IRAF is distributed by the National Optical 
Astronomy Observatories, which are operated by the Association 
of Universities for Research in Astronomy, Inc., under 
cooperative agreement with the National Science Foundation.}
with a 4 pixel radius aperture and an
annulus for sky subtraction with inner and outer radii of 7 and 11 pixels,
respectively.  
After doing an aperture correction as in \citet{holtzman95b} and the
WFPC2 Data Handbook, the F300W, F380W, F555W, F606W, and F814W
magnitudes 
are 17.5, 18.6, 18.8, 18.7, and 19.2, respectively.
Iteratively using
the conversions in the IRAF {\it synphot}
database,
we find V $\approx$ 18.8, U $-$ B $\approx$ $-$0.6,
B $-$ V $\approx$ $-$0.1, and V $-$ I $\approx$ $-$0.4.

As discussed in Section 4.1, the extinction to the ionized gas
in the nuclear region is A$_V$ $\approx$ 1 $-$ 2.  However,
from the slope of the 
UV spectrum, \citet{gonzalez99} conclude that 
the stars in the nuclear
region that are contributing to the UV flux
are significantly less obscured on average than the ionized gas, 
A$_V$ $\approx$ 0.1.
A low extinction 
is consistent with the very blue optical colors of this source, 
typical of B stars.
The absolute V magnitude
of this optical source is therefore $\approx$$-$14.1 or brighter.
This magnitude and the optical diameter of $\le$30 pc
are consistent with those of the brightest
of the so-called `super star clusters' often found
in HST images of interacting galaxies,
which are believed to be very young globular clusters 
\citep{whitmore93, whitmore95,
whitmore99, zepf99,
holtzman92, holtzman96, elmegreen01}.
This source is 
bluer than most of the super star clusters found to date
\citep{whitmore99, zepf99, elmegreen01}, with colors that imply
a cluster age of $<$4 Myrs \citep{leitherer99}.

This optical source is just one of more than 20 such optical clusters
seen in the inner region of NGC 7714 (see Figure 7b), thus it is
possible it is just a chance alignment of an optical cluster with
the ULX.
Assuming the association is correct,
this source has an apparent X-ray-optical ratio of 
log(F$_X$/F$_V$)
$\approx$ $-$0.3 (calculated as in \citealp{maccacaro88}).
This source is somewhat bluer and more optically-luminous
than the possible optical counterparts to about a dozen ULXs
found in the Antennae galaxy 
(M$_V$ = $-$10 to $-$13.7,
B $-V$ = 0 to 0.8, 
log(F$_X$/F$_V$) = $-$1.5 to 0.7;
\citealp{zezas02b}).
Excluding the Antennae sources,
most optical counterparts to ULXs found so far are 
much less optically-luminous than the NGC 7714 source, but similarly blue.
In Table 4, we tabulate information about other possible optical counterparts
to ULXs from the literature.
The NGC 1637, M81, M82 Einstein \#11, M83, and NGC 5204 ULXs
have similar optical colors, 
but much lower optical luminosities \citep{stocke91, immler99, immler03,
liu02, liu04}, while the NGC 4565 source and the M82 MGG-11 source
are both redder and less
luminous \citep{wu02}.
Most of these other ULXs have optical magnitudes and colors consistent with single
O and B stars,
thus they may be compact objects in accreting binary systems.
In contrast, the NGC 7714 and Antennae ULX optical counterparts are consistent with 
very luminous young star clusters.  Perhaps these sources
are accreting binary systems within or recently ejected
from very young star clusters.

The optical magnitude of the apparent optical counterpart to
the secondary nuclear source in NGC 7714 is similar to the nuclei
of the `normal' non-AGN spirals studied by \citet{phillips96}
and \citet{boker04}, who suggest
that these nuclei are simply compact star clusters.
This source is about 5 magnitudes less luminous in V
than typical Seyfert nuclei \citep{chen85, granato93}, thus it does
not appear to be a second obscured nucleus.
The presence of variability
argues that the majority of the 
X-ray emission is likely associated
with a single source, however,
the timescales
involved do not strongly limit the mass of
the emitting object.  The observed
variability could be consistent with either
a stellar mass or supermassive black hole.

\subsection{The Superwind }

\subsubsection{The Age of the Superwind}

The extended X-ray emission in the central region of NGC 7714 may be due at
least in part to a superwind from the nucleus.
The 1.9 kpc radial extent of this emission
can be used to estimate the age
of the superwind, assuming it is a `superbubble' of hot gas produced
by the combined effect of supernova and stellar winds from
the nuclear starburst.
The expansion timescale of a superbubble can be estimated from
$ t_7^{-3/5} \approx 2.8 {\eta}^{1/5}L_{mech}^{1/5} n_H^{-1/5} / r_{shell} $,
where $t_7$ is the timescale in units of 10$^7$ years, L$_{mech}$ is the
mechanical energy of the supernovae in units of 10$^{43}$ erg~s$^{-1}$,
n$_H$
is the average hydrogen number density in units of 1~cm$^{-3}$, 
$\eta$ $\approx$ 0.1 is the efficiency
of the kinetic energy deposited into the gas \citep{dw80},
and r$_{shell}$
is the radius of the bubble in kpc
\citep{shull95,taniguchi01}.
Using
L$_{mech}$ $\approx$ E$_{SN}$R$_{SN}$, 
E$_{SN}$ $\approx$ 10$^{51}$ erg
$\approx$ the energy of a single supernova, 
and hydrogen number density n$_H$ $\approx$ 1 cm$^{-3}$
gives L$_{mech}$ $\approx$ 2.3~$\times$~10$^{41}$~erg~s$^{-1}$, using
a supernova rate R$_{SN}$ of 0.07~yr$^{-1}$.
This implies a timescale of $\approx$15 Myrs
for the bubble.  Allowing time for supernova
activity to begin, this timescale is consistent with the age inferred
for the possible intermediate-age starburst
in NGC 7714 \citep{lancon01}, which is likely the dominant source
of the supernovae which power this wind (Section 4.2).

\subsubsection{The Hard Component of the Extended Emission}

The extended emission
seen in \chandra\ observations
of many starburst galaxies 
can be fit to temperatures ranging from 0.1 $-$ 0.9 keV,
without needing a hard component
(e.g., NGC 4631, \citealp{wang01};
NGC 4676, \citealp{read03}; NGC 4038/9,
\citealp{fabbiano03,fabbiano04}; NGC 4449,
\citealp{summers03}; NGC 1800, \citealp{rasmussen04}).
In contrast, spectral fitting of the extended emission
in NGC 7714 requires
a second hard component
in
addition to a colder thermal component.
NGC 7714 is not unique, however.  
In M82, a 2 $-$ 5 keV
diffuse component has been detected \citep{griffith00, strickland04a}, while
in NGC 3256 three thermal components 
at 0.6, 0.9, and 
3.9 keV 
were fit to the extended
emission
\citep{lira02}.
In some cases, in addition to one or two 0.1 $-$ 0.9 keV thermal
components,
a hard power law component was also included in the fit
(NGC 4214, \citealp{hartwell04};
NGC 5253, \citealp{summers04}; M83, \citealp{soria02}).

A hard spatially-extended
component in galaxies is often assumed to be due to unresolved
X-ray binaries, either low mass (LMXBs), which are associated with
the older stellar population, or the HMXBs discussed in Section 4.2.
An upper limit to the X-ray luminosity from LMXBs 
can be 
obtained from the lower envelope of 
the global L$_B$-vs-L$_X$ plot for elliptical and S0 galaxies
\citep{canizares87, ciotti91}.
In X-ray faint, gas-poor early-type galaxies without much on-going
star formation, the total X-ray luminosity is
an upper limit to the emission from LMXBs.
This limit is L$_X$(LMXB) $\le$ 1.5 $\times$ 
10$^{40}$L$_B$/(10$^{11}$ L$_{\sun}$) erg~s$^{-1}$
\citep{ciotti91}, where L$_B$ is defined as L$_B$ = dex[$-$0.4(M$_B$ $-$
5.41)] \citep{canizares87}. 
Since 
\chandra\ observations of some of
these X-ray-faint early type galaxies show that at most $\approx$30$\%$
of this flux can be attributed to unresolved LMXB
with L$_X$ $\le$ 10$^{38}$ erg~s$^{-1}$
\citep{sarazin01,
sivakoff03, osullivan04},
this limit can be lowered by a factor of $\approx$3.
These conversions
are somewhat uncertain, as the L$_X$(LMXB)/L$_B$ ratio for galaxian
bulges may vary from galaxy to galaxy \citep{irwin02}.

The uncorrected blue luminosity of NGC 7714 
within a 21$''$ aperture
is
6.6 $\times$ 10$^9$ L$_{\sun}$ 
\citep{huchra77},
approximately the area of the superwind.
This is an over-estimate of the light due to
old stars, 
since it contains
the starburst nucleus as well as HII region A.
Using A$_V$ $\approx$ 1.5
as an upper limit to the extinction to the older stars
in the disk
(see Sections 4.1 and 4.3) 
gives L$_X$(LMXB) $\le$ 2 $\times$
10$^{39}$ erg~s$^{-1}$, $\approx$10$\%$ of the extended hard X-ray component. 
Thus LMXBs likely do not account for all of this hard radiation.

An estimate of the contribution 
by HMXBs to the extended X-ray emission in the NGC 7714 disk
can be
made from the H$\alpha$ emission in the `superwind' region and the
\citet{helfand01} L$_X$(HMXB)/O star ratio for Local Group galaxies.
The total extinction-corrected
H$\alpha$ luminosity of NGC 7714 minus
the HII regions and the inner $\approx$1$''$ is $\approx$ 2.8 $\times$ 10$^{41}$ 
erg~s$^{-1}$
\citep{gonzalez95, gonzalez99}.
This implies L$_X$(HMXB) $\le$ 1.0 $\times$ 10$^{40}$ erg~s$^{-1}$,
$\approx$50$\%$ of the hard X-ray luminosity of this region.
Thus HMXBs may be responsible for at least part of this 
hard emission.

Another possible source of extended hard X-rays in a starburst
galaxy is 
inverse Compton radiation caused by the interaction
of infrared photons with supernova-generated relativistic electrons,
as suggested for the starburst galaxies M~82 and NGC 3256 
\citep{rieke80, schaaf89, moran97, moran99}.
In this scenario, the hard X-ray photon
index $\Gamma$ should be equal to $\alpha + 1$, where $\alpha$
is the radio spectral index (F$_{\nu}$ $\propto$ $\nu$$^{-{\alpha}}$).
At high spatial resolution, the radio continuum emission in the
core of NGC 
7714 is extended over $\sim$6$''$, with a 5 GHz flux density
of $\approx$17 mJy and a 5 GHz/1.4 GHz spectral index of $\alpha$
$\approx$ 0.6 \citep{weedman81}.   This is consistent with
the Chandra photon index of $\Gamma$ = 1.8 $\pm$ 0.2
for the hard component of the extended emission.
An exact calculation of the expected X-ray flux from the inverse 
Compton process depends upon the strength and geometry
of the electron, photon, and magnetic fields, which are highly
uncertain.  An approximate estimate can be made from the radio
continuum data, scaling from M~82 and NGC 3256.
For both M~82 and NGC 3256, the ratio of the 5 keV flux
density to the core 5 GHz flux density $R_{5,5}$ $\approx$
10$^{-18}$ erg~s$^{-1}$~cm$^{-2}$~keV$^{-1}$~$\mu$Jy$^{-1}$ 
\citep{moran99}.
For NGC 7714, the 5 keV flux density for the extended emission
is $\approx$1.4 $\times$ 10$^{-14}$ erg~s$^{-1}$~cm$^{-2}$~keV$^{-1}$,
giving a consistent value of $R_{5,5}$ = 
0.8 $\times$
10$^{-18}$ erg~s$^{-1}$~cm$^{-2}$~keV$^{-1}$~$\mu$Jy$^{-1}$.
Thus it is possible that inverse Compton may contribute in NGC 7714
as well.  

\subsection{Extended X-ray Emission from H~II Region Complexes }

\subsubsection{Contributions from HMXBs}

We have detected
extended X-ray emission from four extra-nuclear H~II region
complexes in NGC 7714.
In Table 3, we compare the 0.3 $-$ 8 keV X-ray 
luminosities 
for these HII region complexes with 
the number of Lyman continuum photons N$_{LyC}$ and  
H$\alpha$ luminosities, after correction for an average 
extinction of A$_V$ = 0.7 
\citep{gonzalez95}.
As in the nuclear region, the two most likely contributors to
the extended X-ray emission from these regions are HMXBs and
hot gas produced by
stellar winds and supernovae impacting 
the ambient interstellar medium. 
The Local Group N$_{LyC}$/L$_X$(HMXB) ratios 
are $\approx$3 $-$ 60 $\times$ 10$^{13}$ \citep{helfand01},
thus, except for region D, the X-ray luminosities 
for these HII regions 
are larger than expected for HMXBs.  This indicates
either other contributions to the X-ray flux or an L$_X$(HMXB)/N$_{LyC}$
ratio higher than the galaxian averages tabulated in \citet{helfand01}.

\subsubsection{Contributions from Wind-Heated Gas}

To estimate the contributions to the extended X-ray
emission of these HII regions from wind-heated gas, we use
theoretical models 
of the mechanical luminosity from supernovae and stellar winds
in star clusters from
the {\bf Starburst99}\footnote{Available 
at http://www.stsci.edu/science/starburst99.}
spectral synthesis software, version 4.0 \citep{leitherer99}.
This version
includes improvements to supernova-related quantities
by M. Cervino and updated W-R and O star model atmospheres
by 
\citet{smithr02}.
To determine the X-ray luminosity, we need to assume an X-ray production
factor L$_X$/L$_{mech}$, the fraction of the mechanical energy
of the starburst that is converted into X-rays.  Estimates of 
L$_X$/L$_{mech}$ in the literature vary widely, from
5$\%$ 
\citep{strickland99} to 0.2$\%$ \citep{strickland04b} to 0.02$\%$
\citep{dorland87}.
We discuss 
observational limits on 
L$_X$/L$_{mech}$
further in Section 4.5.3.

In Figure 12, we plot the expected 
extinction-corrected
N$_{LyC}$/L$_X$ 
and 
L$_{H\alpha}$/L$_X$ ratios
from {\bf Starburst99}
for HII regions as a function of star cluster age, including
only contributions to the X-ray flux from mechanical energy
input from stellar winds and supernovae, excluding point sources.
We plot the expected theoretical values
for eight different models in Figure 12.
We include models with continuous star formation as well as 
models with an instantaneous burst of star formation.  We include models
with a Salpeter initial mass function 
(power law index $\alpha$ = 2.35) as well as models with 
$\alpha$ = 3.3.  We have also varied the metallicity from solar metallicity
to 0.25 solar and L$_X$/L$_{mech}$ from 5$\%$ to 0.02$\%$.
Since supernova do not occur until a burst is $\approx$3.6 Myr old,
the model X-ray luminosity before 3.6 Myr
is due solely to stellar winds.
After 6 Myr, the 
mechanical energy from supernova strongly dominates
the production of X-rays, and between 3.6 and 6 Myr, both
winds and supernovae contribute.

The NGC 7714 HII regions 
have N$_{LyC}$/L$_X$ and 
L$_{H\alpha}$/L$_X$ ratios
consistent
with the {\bf Starburst99} models if the ages are between 3.8 $-$ 7 Myrs
(instantaneous burst) or $\ge$5 Myrs 
(if the star formation has been continuous) 
and if the X-ray production efficiency L$_X$/L$_{mech}$ $\approx$ 5$\%$.
These age estimates are consistent with those determined 
independently 
by \citet{garcia97} 
using optical emission line strengths, photoionization
modeling, and evolutionary synthesis models.  For HII regions A, B, and C,
\citet{garcia97}
found ages of 5.0 $\pm$ 0.5 Myr, 3.5 $\pm$ 0.5 Myr, and 
4.5 $\pm$ 0.5 Myr and metallicities of 
0.4, 0.2, and 0.2 solar, respectively, assuming a Salpeter IMF.
Using the \citet{garcia97} ages,
we have included points for these HII regions in 
Figure 12.  This consistency with the {\bf Starburst99}
predictions suggests that mechanical energy from
supernovae does contribute significantly to the X-ray flux
from these HII regions, if the X-ray production efficiency is high.
We discuss L$_X$/L$_{mech}$ further in Section 4.5.3.

The low 
L$_X$/L$_{mech}$ ($\approx$0.02$\%$) instantaneous burst models also give 
L$_X$/L$_{H\alpha}$ ratios consistent with those observed, but only  
if the age is $\approx$15 Myrs, much larger than
the \citet{garcia97} estimates.
This implies, however, that when the HII regions were
younger, their 
L$_{H\alpha}$ would have been unreasonably 
large (L$_{H\alpha}$ $\approx$ 10$^{43}$ erg~s$^{-1}$
at 5 Myrs).
Thus if the X-ray production efficiency is low,
supernovae heating is likely not the dominant source
of X-ray production in these HII regions.

For completeness,
in Table 3 we also include the extinction-corrected
N$_{LyC}$/L$_X$ and L$_{H\alpha}$/L$_X$ ratios for the nucleus.
These values are a factor of 3 $-$ 36 times lower than those
of the HII regions.
In Figure 12,
we have plotted the location
of the NGC 7714 nucleus, using the starburst age of 5 Myr
\citep{gonzalez99,
lancon01}.
As noted in Section 4.2 and discussed by \citet{lancon01},
N$_{LyC}$/L$_X$ for the nucleus
is
approximately a factor of 10 too low compared
to that expected for a 5 Myr starburst
(assuming a 5$\%$ X-ray production
efficiency).  This supports the idea that either there is 
an intermediate-age population present 
which boosts the supernova rate, or 
there are X-ray-bright
point source(s) contributing to the X-ray flux.

Table 3 and Figures 5 and 6 show that 
HII region E has an N$_{LyC}$/L$_X$ ratio a factor 
of 10 lower than that of the other NGC 7714 extra-nuclear HII regions.
It is unclear at present if this difference is caused
by a difference in age or in extinction, since 
no direct estimate of the age or extinction of region E is available
at present.

\subsubsection{Comparison to Other HII Regions and the X-Ray Production
Factor L$_X$/L$_{mech}$}

At the present time, L$_X$/L$_{mech}$ is not well-determined theoretically.
Standard models of HII regions and superbubbles
predict too many X-rays in young HII regions with stellar wind-driven bubbles
\citep{rauw02, dunne03, townsley03, townsley04} and too few
in older supernovae-driven bubbles (e.g., \citealp{chu90, wang91, oey96}).
In this section, we provide observational constraints
on 
L$_X$/L$_{mech}$ by 
compiling X-ray 
luminosities
for some well-studied Galactic
and extragalactic 
HII regions and `superbubbles' from the literature (Table 5) and comparing
to {\bf Starburst99} models and other estimates of L$_{mech}$.
In Table 5, 
we also include
published estimates of the number of Lyman continuum
photons, N$_{LyC}$, and 
extinction-corrected
H$\alpha$ luminosities, based on radio, infrared, and/or
optical observations.
When available,
Table 5 also contains
information about the age of the associated 
star cluster from color-magnitude diagrams
(note that in several cases
more than one cluster with different ages is present in a single HII 
region complex).
In Figure 12,
we have plotted the
HII regions in Table 5 that have estimated cluster ages with the 
{\bf Starburst99} models.

Although there are large uncertainties on
the values in this Table because of the extinction corrections,
the different X-ray energy ranges used,
and assumptions made about the X-ray spectrum,
a general trend is apparent in Table 5 and Figure 12.
The very young HII regions ($<$3.6 Myrs; with no supernovae activity)
have very high
N$_{LyC}$/L$_{X}$ and
L$_{H\alpha}$/L$_{X}$ ratios compared to the older star formation regions.
For these young HII regions,
the 
N$_{LyC}$/L$_X$ and
L$_{H\alpha}$/L$_X$ ratios 
are not consistent with 
the {\bf Starburst99} predictions unless the X-ray production efficiency
L$_X$/L$_{mech}$ is very low. 
Comparing the L$_X$ values 
for the $<$3.5 Myr old HII regions
in Table 5 with
solar metallicity Salpeter IMF
{\bf Starburst99} models gives X-ray production efficiencies L$_X$/L$_{mech}$
that range from 7 $\times$ 10$^{-5}$ to 4 $\times$ 10$^{-3}$, with a median
of 2 $\times$ 10$^{-4}$ (0.02\%).
For some of the young HII regions in Table 5, it is also possible to calculate 
L$_{mech}$ more directly, by summing over the mass loss rates \.M and 
the
terminal wind velocities $V$ of the individual stars in the cluster:
L$_{mech}$ = $\Sigma$$1\over{2}$\.M$_iV_i^2$.   Utilizing mass loss rates
and terminal velocities 
from \citet{stevens03} for the HII regions in our sample, 
we obtained a similar
median L$_X$/L$_{mech}$ of 10$^{-4}$ (0.01\%).

The very weak diffuse X-ray emission
in young HII regions has been
noted before \citep{rauw02, dunne03, townsley03, townsley04}, 
and is not well understood.  Some suggested explanations 
for the low efficiency of forming X-rays
in wind-driven bubbles
include
the escape of hot gas through blowouts and fissures in the interstellar
medium,
suppression of heat conduction by strong magnetic fields, 
the
entrainment of cold material in the wind, and the dissipation
of energy by turbulence.

On average,
the older star formation regions in Table 5
have higher X-ray
production efficiencies than the younger ones.
Comparison with the {\bf Starburst99} instantaneous burst 0.25 solar metallicity
Salpeter
models for the 
$>$3.5 Myr-old LMC superbubbles in Table 5 gives 
a median L$_{X}$/L$_{mech}$ of 2 $\times$ 10$^{-3}$
(0.2$\%$), with a large scatter (0.01 $-$ 5$\%$).
Thus a larger fraction of the total wind
energy gets converted into X-rays in bubbles powered by supernovae than
in stellar wind-driven bubbles.
With a continuous burst, the median L$_X$/L$_{mech}$ is even higher, 7$\%$, with
a range of 0.1 $-$ 36$\%$.
For three of the superbubbles in Table 5, \citet{oey96} has
calculated the mechanical luminosity directly by adding up contributions
from the individual stars in the corresponding
OB associations, and accounting for 
stars that have already become supernovae.  Comparing her L$_{mech}$ values
with
the X-ray luminosities for the affliated superbubbles, we find
L$_X$/L$_{mech}$ $\approx$ 0.02 $-$ 0.05 (2 $-$ 5$\%$), 
higher on average than 
the {\bf Starburst99} results but overlapping in range.

Given the scatter in the empirically-determined X-ray production factors,
it is possible that supernovae-driven winds are responsible for
the observed X-rays in the NGC 7714 HII regions.  
However, it is clear from Figure 12
that the NGC 7714 
HII regions have an X-ray excess compared to LMC superbubbles with
similar ages.
Thus if hot gas is responsible for the extended X-ray emission
in the NGC 7714 HII regions, the X-ray production efficiency must be 
higher on average than in the tabulated LMC regions.
Given the uncertainty in L$_X$/L$_{mech}$,
whether hot supernovae-heated gas or HMXBs are responsible
for the extended X-ray emission from these HII regions is still
an open question.
Further investigations are 
needed to better determine L$_X$/L$_{mech}$ for wind-driven X-ray production.

\subsubsection{Mimicking of ULXs in High Redshift
Galaxies }

In galaxies more distant than NGC 7714, 
HII region complexes similar to those in NGC 7714 may be unresolved
with \textsl{Chandra}, mimicking ULXs.  For example, in the Cartwheel galaxy,
which is three times the distance of NGC 7714, more than a dozen
unresolved
($\le$0.3 $-$ 1.5 kpc)
X-ray sources are detected in the star forming ring, with luminosities
in the ULX range \citep{gao03, wolter04}.   
The sizes and X-ray luminosities of these sources are consistent with
the HII regions seen in NGC 7714, while
H$\alpha$ luminosities from J. Higdon (2004, private communication) give
L$_{H\alpha}$/L$_X$ ratios of $\sim$0.3 $-$ 24,
similar to those 
of the HII regions in NGC 7714.  
Thus at least some of these Cartwheel sources may be similar to the NGC 7714
HII regions, and may be resolved at higher spatial resolution.
Most of the ULXs found so far, however, are in closer galaxies, and thus
are clearly a different class, or classes, of objects.  

\subsection{The Candidate Ultraluminous X-ray Sources }

The luminosity functions of 
ULXs in galaxies appear to scale with star formation rate 
\citep{grimm03, humphrey03, schwarz04},
or perhaps with a combination of star formation rate and galaxy
mass.
Furthermore, ULX candidates tend to be positionally coincident
with star formation regions, for example, most of 
the ULXs in the interacting galaxy M51 are associated with spiral arms
\citep{terashima04}.
This correlation with star formation
has been used to argue for the beamed model of these sources
\citep{king01} or alternatively, for an intermediate mass black hole
scenario where accretion from molecular clouds is occuring 
(e.g., \citealp{krolik04}).  

In NGC 7714/5, only two of the candidate ULX sources in Table 1
are associated with on-going star formation, \#8 and \#9.
Interestingly, of the 11 ULX candidates, eight appear to be
in or near tidal features or the ring (\#1, \#4, \#5,
\#8, \#9, \#10, \#11, and \#12)
and two are near the ends of the NGC 7714 bar
(\#6 and \#7),
which may
also have been produced by the interaction. 
One of the remaining sources,
\#2, lies near the center of 
NGC 7715,
which is believed to be in a post-starburst state with a stellar
population age of
$\sim$45 Myr \citep{bernlohr93}.
This lack of association with star formation regions
does not
rule out the beamed stellar mass black hole scenario, however.
In the high mass
X-ray binary population synthesis models of 
\citet{rappaport04}, the ULX population is largest 5 $-$ 100 Myrs
after a burst of star formation.  In their scenario, a larger ULX
population is expected in post-starburst regions 
(or regions with
on-going star formation) than in very young bursts.
In NGC 7714,
it is possible that the tails and rings 
are post-starburst,
since starbursts can be triggered in such features
\citep[e.g.,][]{schombert90}. 
Near-infrared colors suggest that the ring may have older
stellar population \citep{bushouse90},
but at the present time 
no information about population ages
in the NGC 7714 tails is available.
The closest approach between the two galaxies occurred $\approx$170 Myrs ago
\citep{struck03}, 
thus it is possible that the formation of the ULXs in the tidal
features, the ring, and the bar was triggered by the interaction. 

The most luminous point source in the \chandra\ map, \#4,
near the NGC 7714 ring,
was previously detected in two separate XMM-Newton observations 
on June 2002 and December 2002
\citep{soria04}.
The XMM-Newton luminosity increased by a factor of two between
these two observations, with a 0.3 $-$ 12 keV luminosity 
of 6.6 $\times$ 10$^{40}$ erg~s$^{-1}$ in December 2002.
In the high-luminosity state, the XMM-Newton spectrum can be well-fit by
a single $\Gamma$ = 2.1 $\pm$ $^{0.2}_{0.1}$ power law.
In the low-luminosity 
state, in addition to a $\Gamma$ = 2.6 $\pm$ $^{0.4}_{0.7}$
power law, a second component was required.  This second component
could be represented by a 1 keV blackbody.
The \chandra\ luminosity,
L(0.3 $-$ 8 keV) = 6.4 $\times$ 10$^{40}$ erg~s$^{-1}$, extrapolates to
L(0.3 $-$ 12 keV) = 8.0 $\times$ 10$^{40}$ erg~s$^{-1}$, which
is similar to that in the 
XMM-Newton
high-luminosity state, 
and the power law index, $\Gamma$ = 1.9 $\pm$ $^{0.1}_{0.0}$,
is consistent within the uncertainties.
Thus the \chandra\ observations were also made during
a high-luminosity state.
The fact that this source is point-like in the \chandra\
data and variable in XMM-Newton rules out the possibility
that the X-rays are arising from hot gas heated by
the impact of 
bridge material or nuclear outflow, 
as suggested by 
\citet{papaderos98}
based on low spatial resolution data from ROSAT.

As noted earlier, source \#12 has undergone a factor of 10 increase
compared to previous XMM observations.  This argues for a single point-like
source, although does not distinguish between a low mass or supermassive
black hole.  If the 19 second flare is real, however, this constrains
the system mass to $^{<}_{\sim}$4 $\times$ 10$^4$ M$_{\sun}$.  

The \chandra\ column density for source
\#4, 1.9 $\pm$ 
0.3
$\times$
10$^{21}$ cm$^{-2}$, is consistent with that found from XMM-Newton
and with the HI column density from \citet{smith97}.
In contrast,
for source \#12, 
which is near the edge of the outer tail,
the low column density is consistent with 
the Galactic foreground.

\section{Summary}

We have obtained \chandra\ X-ray images of the prototypical nuclear
starburst galaxy NGC 7714 and its companion NGC 7715.  
Our primary results are:

1. The 0.3 $-$ 8 keV luminosity of the inner 150 pc radius
nuclear region of NGC 7714
is $\approx$4 $\times$ 10$^{40}$ erg~s$^{-1}$.  
This is high
compared to the number of Lyman continuum photons,
suggesting that it is due to an enhanced population of HMXBs,
one or more
ULXs, an obscured AGN, or 
hot gas efficiently heated by supernovae produced by an 
intermediate-age
stellar
population.
The possible variability in the Chandra data suggests that 
a single source contributes significantly to the observed X-ray flux,
however, the timescale of variation and the lack of Fe-K line emission
argue against an AGN.

2. A second luminous (L$_X$ $\approx$ 10$^{40}$ erg~s$^{-1}$)
point source is detected 1\farcs5 (270 pc) northwest of the nucleus,
with an apparent optical counterpart on HST images.
The optical magnitude of this source (M$_V$ $\approx$ $-$14.1) and
its blue optical colors (B $-$ V $\approx$ $-$0.1) are consistent
with it being 
a very young ($<$4 Myr) globular cluster. 
This source is much more optically-luminous than most optical
counterparts to ULXs 
found
to date, however, it has similar optical colors.
This suggests that it may be a mass-transfer
X-ray binary with a young massive stellar companion, which resides 
in a young star cluster.
The observed \chandra\ variability is consistent with this suggestion.

3. Diffuse emission extending 11$''$ (1.9 kpc) to
the north of the nucleus is visible
in the \chandra\ images.  
The X-ray spectrum can be fit by two
Mekal functions, with temperatures of 0.59 $\pm$ $^{0.05}_{0.06}$
keV
and 8 $\pm$ $^{10}_3$ keV, or with a 0.6 keV Mekal function plus
a $\Gamma$ = 1.9 $\pm$ $^{0.3}_{0.0}$ power law.
The hard component contributes
$\sim$2/3 of the 0.3 $-$ 8 keV luminosity.  Much of this hard
component may be due to HMXBs, however, inverse Compton radiation
cannot be ruled out.  
The soft component is probably caused by a superwind powered
by the nuclear starburst.  
The superbubble expansion timescale is $\sim$15 Myr, consistent
with the age of the
intermediate-age stellar population suggested by \citet{lancon01}.
There is some evidence for an enhanced silicon abundance, especially
in the innermost regions.  The presence of an abundance gradient is
less clear.

4. We have detected extended X-ray emission from four extra-nuclear HII
region complexes. The high X-ray luminosities 
relative to 
the
extinction-corrected H$\alpha$ luminosities imply that
either 
an enhanced population of HMXBs or
shock heating
from supernovae with a high X-ray production efficiency
($\approx$5$\%$)
is responsible for the emission.
The high X-ray fluxes of
these star formation regions imply that they
are at least 3.5 Myr old, so that supernovae
and HMXBs
have started to occur.
These ages are consistent with previous estimates based on optical
spectroscopy and photoionization modeling.

5. To better determine the X-ray production efficiency,
we have compiled data on well-studied star formation regions 
in nearby galaxies.  The median X-ray production efficiency
is 0.02$\%$ for 
young ($<$3.5 Myrs) HII regions and 0.2 $-$ 7$\%$ for older
star formation regions.  Thus it is possible 
that the extended emission in the NGC 7714 HII regions is due
to hot gas, however, this is uncertain.

6. In addition to the luminous X-ray source near the nucleus, 
11 other X-ray point sources are visible in the vicinity of NGC 7714/5.
Ten of these appear to be
associated with features likely produced in the interaction, but only two
are in star formation regions.
One of these sources shows evidence of short timescale (19 second)
variability.

7.  No X-ray emission was detected from SN 1999dn or the 
possible supernova noted by \citet{mattila02},
with upper limits to the 0.3 $-$ 8 keV luminosities 
of $\approx$2 $\times$ 10$^{38}$ erg~s$^{-1}$.





\acknowledgments

We thank the \chandra\ team for making this research possible.
We are grateful to 
J. M. Houchins for computer support, and Mark Giroux, Roberto Soria,
Bob Benjamin,
Henry
Kobulnicky, and an anonymous referee for helpful
suggestions.
This research was supported by NASA \chandra\ grant G04-5103A, NSF
grant AST-0097616, and NASA grant SV3-73016 (MAN).
This research has made use of the NASA/IPAC Extragalactic Database (NED) which is operated by the Jet Propulsion Laboratory, California Institute of Technology, under contract with the National Aeronautics and Space Administration.

\clearpage



\figcaption{
  \small A 4\farcm5 $\times$ 2\farcm2 optical photograph of NGC 7714/5
  from \citet{arp66}.  North is up and east is to the left.
  NGC 7714 is the larger galaxy to the west.  Note the partial ring
  (1), the two western tails (2, 3), the northeastern tail of NGC 7714
  (4), the bridge (5), and the eastern tail of NGC 7715 (6).}

\figcaption{
\small A 21 cm {\sc Hi} map of NGC 7714/5 (greyscale,
from \citealp{smith97}), with superposed optical contours
from the smoothed Digitized Sky Survey image.
Note the large {\sc Hi} loop to the west, coincident
with the optical tail, as well as the offset between the gas and the
stars in the bridge.}

\figcaption{
The smoothed \chandra\ maps of NGC 7714/5 in various energy ranges,
with optical Digitized Sky Survey contours.  Top: 0.3 - 8 keV,
created using a weighted exposure map.  
Second from top:
0.3 $-$ 1 keV,
created using a monochromatic 0.8 keV exposure map.
Third from top:
1 $-$ 2.5 keV,
created using a monochromatic 1.5 keV exposure map.
Bottom: 2.5 $-$ 8 keV, 
created using a monochromatic 3 keV exposure map.
}

\figcaption{
The smoothed 0.3 $-$ 8 keV \chandra\ map of NGC 7714/5 (greyscale), with
H$\alpha$ contours from Smith et al. (1997) superimposed.
}

\figcaption{
A zoomed-in view of the NGC 7714 disk, 
with an overlay of the \chandra\ 0.3 $-$ 8 keV
X-ray map
as contours on the Smith et al. (1997) H$\alpha$ map (greyscale).
Extended X-ray emission is seen associated
with HII region complex A, B, C, and E, 
but not with region D, in spite of the fact that D is brighter in H$\alpha$
than E.
Note that the ULX candidate to the southwest of the nucleus is
offset from the middle of HII region C.  Note also
the diffuse X-ray emission extending 11$''$ to the north
of the nucleus, with a possible faint H$\alpha$ counterpart.
}

\figcaption{
An overlay of the \chandra\ 0.3 $-$ 8 keV map (contours)
on an archival HST F606W WFPC2 image of NGC 7714, after registering
the HST map 
so that the brightest X-ray source is coincident 
with the brightest optical source. Note the X-ray 
counterparts to the HII regions to the northwest, 
northwest, and southeast of the nucleus. There are 
no obvious optical counterparts to the four candidate 
ULX sources in the field.
}

\figcaption{
a) Left:
A close-up view of the inner 10$''$ of the \chandra\ 0.3 $-$ 8 keV map
(both contours and greyscale).
Notice the second point source $\approx$1.5$''$ northwest of the nucleus.
The extended source to the southeast is HII region complex A.
b) Right: The \chandra\ contours for the inner 10$''$, superimposed 
on the archival HST F606W image.  Note the bright optical point
source within 1 \chandra\ pixel of the secondary X-ray source.
}

\figcaption{
The smoothed 0.3 - 8 keV \chandra\ map, with X-ray point sources marked
(see Table 1).
The contours are from the optical Digitized Sky Survey.
These point sources were selected by the CIAO WAVDETECT routine.
Note that sources 4 and 12 are very bright.  Interestingly, source
12 was very faint in the 2002 XMM-Newton observations of Soria and Motch (2004),
with a brightness comparable to the other point sources in the field.
Its X-ray luminosity has increased by a factor of about 10 in the \chandra\
observations.   Sources 8 and 9 are associated with the
inner western tail (and HII regions; see Figure 4).
Sources 10, 11, and 12 may be associated with the outer western tail.
Source 1 appears associated with the western NGC 7715 tail, while
source 2 appears to be in the disk of NGC 7715.  
Sources 3 and 13 are likely background sources.
No X-ray counterpart to the Type Ib SN 1999dn was detected.
}

\figcaption{
Background-subtracted \chandra\ spectra (datapoints), best fit spectral
fits (solid lines), and residuals (dotted lines) for various regions in
the NGC 7714/5 system.  
Top left: nuclear spectrum, with power-law fit; 
Top right: spectrum of secondary source
near nucleus with power-law fit.
Middle left: spectrum for the inner diffuse gas of NGC 7714, excluding the
two nuclear sources, fit with two thermal plasmas.
Middle right: spectrum for the outer diffuse gas of NGC 7714, excluding
HII region A, fit with two thermal plasmas.
Bottom left: ULX candidate \#4, with a power law fit.
Bottom right: ULX candidate \#12, with a power law fit.
See text and Table 2 for more details.
}

\figcaption{
X-ray colors of HII regions A, B, C, and E and ULX candidates \#1 $-$ \#13,
with the soft band being defined as 0.3 $-$ 1 keV, medium 1 $-$ 2 keV,
and hard 2 $-$ 8 keV.  The dotted lines show the locations of absorbed
power laws with photon indices of 4, 3, 2, and 1, respectively, 
from left to right.
The column density ranges from 10$^{20}$ cm$^{-2}$ at the lower left
end of these curves, to 10$^{24}$ cm$^{-2}$ at the far right.
}

\figcaption{
Lightcurves (dotted lines) from the nuclear source (left, 1000 sec
bins) and the secondary nuclear source (right, 2000 second bins).  The
results of a `Bayesian Blocks' analysis (see text) have been
over-plotted for both (solid lines).   Assuming a prior probability for
the number of blocks, $N_b$, is $\propto \gamma^{N_b}$, the nucleus
lightcurve is divided into two blocks for $1-\gamma < 0.995$, with
count rates $(2.72\pm0.09)\times10^{-2}$~cps and
$(2.31\pm0.10)\times10^{-2}$~cps.  The secondary source lightcurve is
divided into two blocks for $\gamma > 4 \times 10^{-7}$, with count
rates $(8.7\pm0.4)\times10^{-3}$~cps and
$(1.5\pm0.1)\times10^{-2}$~cps.
}


\figcaption{
Six model
extinction-corrected 
N$_{LyC}$/L$_X$ and 
L$_{H\alpha}$/L$_X$ ratios
for star formation regions as a function of time, from the
{\bf Starburst99} Version 4.0 stellar population
synthesis code (Leitherer et al. 1999).
As identified by the labels in the upper right, the models
differ in their initial mass functions (power law index $\alpha$ = 2.35
and 3.3), metallicity (solar vs. 0.25 solar), X-ray production
efficiency L$_X$/L$_{mech}$ (5\% vs. 0.02\%) and whether they assume an
instantaneous burst or continuous star formation.
Upper and lower mass limits for the IMF of 100 M$_{\sun}$ and 1 M$_{\sun}$,
respectively,
are assumed for all models.
The filled square
datapoints are measurements from the literature for Galactic, LMC, and SMC
HII regions and superbubbles
(Table 5).
The open circles are values for NGC 7714 HII regions A, B, and C,
using ages from Garc\'ia-Vargas et al. (1997),
which were determined using $\alpha$ = 2.35
and metallicities of 0.4, 0.2, and 0.2 solar, respectively.
The location of the NGC 7714 nucleus is also marked (open square), 
assuming a 5 Myr
burst.
}









\input{BeverlySmith.tab1.tex}

\input{BeverlySmith.tab2.tex}
\input{BeverlySmith.tab3.tex}
\input{BeverlySmith.tab4.tex}
\input{BeverlySmith.tab5.tex}


\end{document}

%% file: BeverlySmith.tab1.tex
%
%
\begin{deluxetable}{lrrrrrrrcccrrrccccrr}
\tabletypesize{\scriptsize}
\def\et#1#2#3{${#1}^{+#2}_{-#3}$}
\tablewidth{0pt}
\tablecaption{Candidate ULXs in the NGC 7714/5 Field\label{tab-1}}
\tablehead{
\colhead{Source} &
\multicolumn{1}{c}{Chandra Name}&
\colhead{Net} & 
\colhead{$L_X$} & 
\colhead{Apparent Location}
\\
&
\multicolumn{1}{c}{(J2000 coordinates)}
& \multicolumn{1}{c}{Counts} 
& \multicolumn{1}{c}{(0.3 $-$ 8 keV)} 
&
\\
&
\multicolumn{1}{c}{}
& \multicolumn{1}{c}{(0.3 $-$ 8 keV)} 
& \multicolumn{1}{c}{(10$^{38}$ erg/s)} 
&
}
\startdata
\#1&CXOU 233623.5+020933&10 $\pm$ 3&3&NGC 7715 tail\\
\#2&CXOU 233622.1+020923&28 $\pm$ 5&7&NGC 7715 disk\\
\#3&CXOU 233619.3+020839&50 $\pm$ 7&12&background?\\
\#4&CXOU 233615.6+020923&2657 $\pm$ 52&640&NGC 7714 ring\\
\#5&CXOU 233615.1+020904&21 $\pm$ 5&5&NGC 7714 ring\\
\#6&CXOU 233614.2+020908&21 $\pm$ 5&6&near end of NGC 7714 bar\\
\#7&CXOU 233613.9+020936&26 $\pm$ 5&7&near end of NGC 7714 bar\\
\#8&CXOU 233613.2+020902&30 $\pm$ 6&8&inner western NGC 7714 tail\\
\#9&CXOU 233612.3+020905&22 $\pm$ 5&6&inner western NGC 7714 tail\\
\#10&CXOU 233611.5+020853&10 $\pm$ 3&3&outer western NGC 7714 tail\\
\#11&CXOU 233610.1+020923&30 $\pm$ 6&8&outer western NGC 7714 tail\\
\#12&CXOU 233610.0+020900&1076 $\pm$ 33&240&outer western NGC 7714 tail\\
\#13&CXOU 233609.1+021007&20 $\pm$ 5&5&background?\\
\enddata
\end{deluxetable}
%

%% file: BeverlySmith.tab2.tex
\begin{deluxetable}{cccccccrcccccc}
\tabletypesize{\scriptsize}
\tablecaption{Spectral Fits \label{tbl-2}}
\tablewidth{0pt}
\tablehead{
\colhead{Source} &
\colhead{Function} &
\colhead{$\chi^2$/DoF} &
\colhead{${\rm n}_{\rm H}$}  & 
\colhead{$\Gamma$} & 
\colhead{kT} &
\colhead{$[$Si$]$}&
\colhead{Net} &
\colhead{L$_X$} &
\\
\colhead{}   &
\colhead{} &
\colhead{} &
\colhead{$\times$ 10$^{21}$}&
\colhead{} &
\colhead{(keV)} &
\colhead{} &
\colhead{Counts} &
\colhead{(0.3$-$8 keV)} &
\\
\colhead{}   &
\colhead{} &
\colhead{} &
\colhead{cm$^{-2}$}  & 
\colhead{} &
\colhead{} &
\colhead{} &
\colhead{(0.3$-$8 keV)} &
\colhead{(10$^{40}$ erg/s)} &
\\
}
\startdata
Nucleus$^{a,b}$&Mekal&60/61&1.5$\pm$$^{0.3}_{0.4}$&&9$\pm$$^{4}_{2}$&
&1498$\pm$39
&4.7\\
Nucleus$^{a}$&Power Law&60/61&2.2$\pm$0.2&1.7$\pm$$^{0.1}_{0.0}$&&
&1498$\pm$39
&4.4\\
Secondary Source&Power Law&17/23&2.0$\pm$$^{0.9}_{0.7}$&&1.9$\pm$$^{0.3}_{0.0}$&
&480$\pm$22
&1.1\\
Central Diffuse$^c$&Mekal+Mekal&59/60&0.5$\pm$$^{0.4}_{0.3}$&&
0.59$\pm$$^{0.05}_{0.06}$/8$\pm$$^{10}_3$&1.2$\pm$$^{0.6}_{0.7}$
&1933$\pm$44
&3.5\\
Central Diffuse$^d$&Mekal+Power Law&59/60&1.0$\pm$$^{0.6}_{0.4}$&
1.8$\pm$0.2&0.57$\pm$$^{0.05}_{0.34}$&0.9$\pm$0.7
&1933$\pm$44
&3.5\\
Inner Diffuse$^e$&Mekal+Mekal&27/40&0.7$\pm$$^{0.7}_{0.5}$&&
0.6$\pm$$^{0.08}_{0.12}$/5$\pm$$^3_1$&4.2$\pm$$^{3.4}_{2.6}$
&1163$\pm$34
&2.4\\
Inner Diffuse$^e$&Mekal+Power Law&28/40&1.4$\pm$$^{3.4}_{0.6}$&
1.9$\pm$$^{0.4}_{0.2}$&0.55$\pm$$^{0.10}_{0.33}$&8.4$\pm$$^{20.5}_{5.5}$
&1163$\pm$34
&2.3\\
Outer Diffuse$^f$&Mekal+Mekal&19/23&0$\pm$$^{5.7}_0$ && 
0.58$\pm$$^{0.05}_{0.07}$/4$\pm$$^6_1$&1.2$\pm$$^{5.6}_{1.2}$
&742$\pm$28
&1.2\\
Outer Diffuse$^f$&Mekal+Power Law&19/23&0.6$\pm$$^{4.4}_{0.5}$&
2.2$\pm$$^{0.6}_{0.5}$& 0.56$\pm$$^{0.07}_{0.23}$& 0.5$\pm$$^{4.9}_{0.5}$
&
742$\pm$28
&1.1\\
Point Source \#4$^a$&Power Law&78/93&1.9$\pm$0.3&1.9$\pm$$^{0.1}_{0.0}$&&&
2657$\pm$52
&
6.4\\
Point Source \#12$^a$&Power Law&51/41&0.6$\pm$0.5&1.6$\pm$0.2&&
&1046$\pm$32
&2.4\\
\\
 \enddata

\tablenotetext{a}{
Used pileup model with $\alpha$ = 0.5 and psffract = 0.95 fixed.
Models without pileup give consistent results.
}
\tablenotetext{b}{
These parameters give acceptable fits for all abundances $<$2 solar.}
\tablenotetext{c}{In a 9\farcs5 region centered 1\farcs5E, 2$''$N of the nucleus,
excluding the two central point sources and HII region A. The metal
abundance was fixed at 0.25 solar.  All abundances
from 0.1 $-$ 2 solar give acceptable fits with consistent parameters.}
\tablenotetext{d}{The metal abundance, aside from $[$Si$]$, 
was fixed at 0.25 solar.  The abundance
could not be constrained.}
\tablenotetext{e}{In a 3$''$ region centered 1\farcs5E, 2$''$N of the nucleus,
excluding the two central point sources.
The abundances of the other elements besides silicon were fixed to 0.25 solar.}
\tablenotetext{f}{In a 3$''$ $-$ 9\farcs5 annulus centered 1\farcs5E, 2$''$N of the nucleus,
excluding the two central point sources and HII region A.
The abundance of the other elements besides silicon was fixed at 0.25 solar.}

\end{deluxetable}

%% file: BeverlySmith.tab3.tex
\begin{deluxetable}{crcrrrrrrrrrrrr}
\tabletypesize{\scriptsize}
\tablecaption{HII Region Complexes in NGC 7714\label{tbl-3}}
\tablewidth{0pt}
\tablehead{
\colhead{Region} &
\colhead{L$_X$} &
\colhead{N$_{LyC}$} &
\colhead{L$_{H\alpha}$} &
\colhead{N$_{LyC}$/L$_X$} &
\colhead{L$_{H\alpha}$/L$_X$} \\
\colhead{} &
\colhead{(erg s$^{-1}$)}   &
\colhead{(photons s$^{-1}$)}   &
\colhead{(erg s$^{-1}$)}   &
\colhead{(photons/erg)}  &
\colhead{} \\
\colhead{} &
\colhead{(0.3 $-$ 8 keV)}   &
\colhead{} &
\colhead{} &
\colhead{}   &
\colhead{} \\
}
\startdata
A&1.3 $\times$ 10$^{39}$&2.0 $\times$ 10$^{52}$&2.8 $\times$ 10$^{40}$&
1.5 $\times$ 10$^{12}$&
22\\
B&4.8 $\times$ 10$^{39}$&3.8 $\times$ 10$^{52}$&5.2 $\times$ 10$^{40}$&
7.9 $\times$ 10$^{12}$&
11\\
C&4.9 $\times$ 10$^{39}$&2.2 $\times$ 10$^{52}$&3.0 $\times$ 10$^{40}$&
4.5 $\times$ 10$^{12}$&
6.1\\
D&$\le$7.3 $\times$ 10$^{38}$&1.2 $\times$ 10$^{52}$&1.7 $\times$ 10$^{40}$
&$\ge$1.6 $\times$ 10$^{13}$&
$\ge$23
\\
E&3.9 $\times$ 10$^{38}$&8.7 $\times$ 10$^{50}$&1.2 $\times$ 10$^{39}$&
2.2 $\times$ 10$^{12}$&
3.1\\
Nucleus&4.4 $\times$ 10$^{40}$&4.7 $\times$ 10$^{52}$&6.4 $\times$ 10$^{40}$&
1.1 $\times$ 10$^{12}$&
1.5\\
\enddata

\end{deluxetable}

%% file: BeverlySmith.tab4.tex
\begin{deluxetable}{ccrrrrrrrrr}
\tabletypesize{\scriptsize}
\tablecaption{Possible Optical Counterparts to ULX Candidates in the Literature\label{tbl-4}}
\tablewidth{0pt}
\tablehead{
\colhead{Galaxy}&
\colhead{ULX}&
\colhead{M$_B$}&
\colhead{M$_V$}&
\colhead{U $-$ B}&
\colhead{B $-$ V}&
\colhead{V $-$ I}&
\colhead{B $-$ I}&
\colhead{log(F$_X$/F$_V$)}&
\colhead{Ref}
\\
}
\startdata
NGC 7714&1.5$''$ NE of nuc&$-$14.2&$-$14.1&$-$0.6&$-$0.1&$-$0.4
&$-$0.5
&$-$0.3&a\\
NGC 1637&\#68&&$-$7.55&&&$-$0.3&&1.13&b\\
M81&X-11&$-$4.3&$-$4.2&&$-$0.1&&&3.0&c\\
M82&MGG-11&&$-$12.5&&&&&&d\\
M82&Einstein \#1&$-$5.5&&&$<$$-$0.3&&&$\ge$500&e\\
M83&H30&&$-$9&&0&&&0.58&f\\
NGC 4565&RXJ1236.2+2558&$-$4.9&&&&&1.1&540&g\\
NGC 5204&U1&&&&&&&&h\\
\enddata
\tablenotetext{a}{This work.}
\tablenotetext{b}{\citet{immler03}.}
\tablenotetext{c}{\citet{liu02}.}
\tablenotetext{d}{X-ray data from ASCA \citep{matsumoto99}.
A possible association with the optical cluster MGG-11 has been noted
by \citet{portegieszwart04}.
MGG-11 has an observed HST NICMOS near-infrared [F160W]
magnitude of 13.1 $\pm$ 0.15 with extinction A$_{F160W}$ = 1.4 $\pm$ 0.5, an  
inferred age of 7 $-$ 12 Myrs,  and an effective spectral type of M4.5I
\citep{mccrady03}.
Using the IRAF {\bf synphot} database and a distance of 3.6 Mpc,
we infer an absolute K magnitude of $-$16.7 for this
cluster.  Using the \citet{koornneef83}
stellar colors, we obtain M$_V$ $\approx$ $-$12.5.
}
\tablenotetext{e}{Seen in Einstein data \citep{stocke91}, but
not in ROSAT \citep{collura94} or Chandra data \citep{griffith00}.}
\tablenotetext{f}{\citet{immler99}, using a distance of 4.5 Mpc 
\citep{thim03}.}
\tablenotetext{g}{\citet{wu02}.}
\tablenotetext{h}{UV-Optical spectra and fluxes
consistent with B0Ib star \citep{liu04}.}
\end{deluxetable}

%% file: BeverlySmith.tab5.tex
\begin{deluxetable}{ccrrrrrrrrrr}
\tabletypesize{\scriptsize}
\rotate
\tablecaption{Global Properties of Galactic and Extragalactic 
HII Region Complexes from the Literature
\label{tbl-5}}
\tablewidth{0pt}
\tablehead{
\colhead{Galaxy} &
\colhead{HII Region/} &
\colhead{Distance} &
\colhead{Age} &
\colhead{Ref} &
\colhead{L$_X$} &
\colhead{Ref} &
\colhead{N$_{LyC}$} &
\colhead{L$_{H\alpha}^{\dagger}$} &
\colhead{Ref} &
\colhead{L$_{H\alpha}^{\dagger}/{L_X}$} &
\colhead{N$_{LyC}$/L$_X$}
\\
\colhead{} &
\colhead{Superbubble} &
\colhead{} &
\colhead{(Myr)} &
\colhead{} &
\colhead{(erg s$^{-1}$)} &
\colhead{} &
\colhead{(photons/s)} &
\colhead{(erg s$^{-1}$)} &
\colhead{} &
\colhead{} &
\colhead{(photons/erg)} 
\\
}
\startdata
Milky Way&Orion&0.5 kpc&$\le$1&a&
$\le$3.8 $\times$ 10$^{33}$&b&8.0 $\times$ 10$^{48}$&1 $\times$ 10$^{37}$&
c&$\ge$2600&$\ge$1.9 $\times$ 10$^{15}$\\
Milky Way&M8=Lagoon&1.4 kpc&1.5&d&6.6 $\times$ 10$^{32}$&e&
2.1 $\times$ 10$^{49}$&
3 $\times$ 10$^{37}$&c&45,000&3.3 $\times$ 10$^{16}$\\
Milky Way&NGC 2244=Rosette&1.4 kpc&2&f&2 $\times$ 10$^{32}$&g&
5.8 $\times$ 10$^{49}$&
8 $\times$ 10$^{37}$&c&
400,000&2.9 $\times$ 10$^{17}$\\
Milky Way&NGC 3603&8.5 kpc&1&h&2 $\times$ 10$^{34}$&i&
1.1 $\times$ 10$^{51}$&
1.5 $\times$ 10$^{39}$&c&100,000&7.2 $\times$ 10$^{16}$\\
Milky Way&M17=Omega&1.6 kpc&1&j&2.5 $\times$ 10$^{33}$&k&
1.8 $\times$ 10$^{50}$&
2.5 $\times$ 10$^{38}$&
l&200,000&7.6 $\times$ 10$^{16}$\\
Milky Way&Carina&1.4 kpc&1 $-$ 6,2&m&4.6 $\times$ 10$^{34}$&n&
4.4 $\times$ 10$^{50}$&
6 $\times$ 10$^{38}$&c&13,000&9.5 $\times$ 10$^{15}$\\
Milky Way&W51&5.5 kpc&0.4 $-$ 2.3&o&4.5 $\times$ 10$^{33}$&p&
1.7 $\times$ 10$^{51}$&
2.4 $\times$ 10$^{39}$&q&53,000&3.9 $\times$ 10$^{16}$\\
Milky Way&RCW 49&2.3 kpc&2 $-$ 3&r&3 $\times$ 10$^{33}$&p&
1.4 $\times$ 10$^{50}$&
1.9 $\times$ 
10$^{38}$&q&63,000&4.6 $\times$ 10$^{16}$\\
LMC&30 Dor&50 kpc&2 $-$ 5&s&2$ -$ 6 $\times$ 10$^{37}$&t&
1.1 $\times$ 10$^{52}$&
1.5 $\times$ 10$^{40}$&c&
300&2.2 $\times$ 10$^{14}$\\
LMC&N11&50 kpc&4 $-$ 5&u&8 $\times$ 10$^{35}$&v&
7.3 $\times$ 10$^{50}$&
1.0 $\times$ 10$^{39}$&w&
1300&9.5 $\times$ 10$^{14}$\\
LMC&DEM 152&50 kpc&$^{>}_{\sim}$10&x&5.4 $\times$ 10$^{35}$&
y&
2.1 $\times$ 10$^{50}$&
2.9 $\times$ 10$^{38}$&x&540&3.9 $\times$ 10$^{14}$\\
LMC&DEM 25&50 kpc&6&z&1.8 $\times$ 10$^{35}$&z&
2.9 $\times$ 10$^{48}$&
4.0 $\times$ 10$^{36}$&aa&
22&1.6 $\times$ 10$^{13}$\\
LMC&DEM 301&50 kpc&4.5&z&5.4 $\times$ 10$^{35}$&z&
2.0 $\times$ 10$^{50}$&
2.7 $\times$ 10$^{38}$&
aa&
500&3.7 $\times$ 10$^{14}$\\
LMC&DEM 50&50 kpc&4.5&z&4.2 $\times$ 10$^{35}$&z&2.2 $\times$ 10$^{49}$
&3.0 $\times$ 10$^{37}$&z&72&5.2 $\times$ 10$^{13}$\\
SMC&NGC 346&59 kpc&3&bb&
1.5 $\times$ 10$^{34}$&cc&5.9 $\times$ 10$^{50}$&8.1 $\times$ 10$^{38}$&
dd&54,000&3.9 $\times$ 10$^{16}$\\
NGC 4303&B/27&16.1 Mpc&&&1.3 $\times$ 10$^{39}$&ee&
1.7 $\times$ 10$^{52}$&
2.4 $\times$ 10$^{40}$&
ff&19&1.4 $\times$ 10$^{13}$\\
NGC 4303&F/69&16.1 Mpc&&&1.1 $\times$ 10$^{39}$&ee&
2.3 $\times$ 10$^{52}$&
3.2 $\times$ 10$^{40}$&ff&
28&2.0 $\times$ 10$^{13}$\\
M101&NGC 5455&7.2 Mpc&&&4.8 $\times$ 10$^{38}$&gg&1.2 $\times$ 10$^{52}$&
1.7 $\times$ 10$^{40}$&hh&35&1.1 $\times$ 10$^{13}$\\
M101&NGC 5461&7.2 Mpc&&&1.1 $\times$ 10$^{38}$&gg,ii&3.7 $\times$ 10$^{52}$&5.1 $\times$ 10$^{40}$&
hh,jj&90&6.6 $\times$ 10$^{13}$\\
M101&NGC 5462&7.2 Mpc&&&1.9 $\times$ 10$^{38}$&gg,ii&1.8 $\times$ 10$^{52}$&2.5 $\times$ 10$^{40}$&jj&130&
9.6 $\times$ 10$^{13}$\\
\enddata
\tablenotetext{\dagger}{Extinction-corrected.}
\tablenotetext{a}{\citet{odell03}.}
\tablenotetext{b}{\citet{den_boggende78}, total X-ray luminosity from ANS, 1 $-$ 8 keV.  This is
an upper limit to the diffuse emission.}
\tablenotetext{c}{\citet{kennicutt84}.}
\tablenotetext{d}{\citet{sung00}.}
\tablenotetext{e}{\citet{rauw02}, with XMM-Newton, 0.5-2 keV.}
\tablenotetext{f}{\citet{park02}.}
\tablenotetext{g}{\citet{townsley03}, Chandra, 0.5 $-$ 2 keV.}
\tablenotetext{h}{\citet{sung04}.}
\tablenotetext{i}{\citet{moffatt02}, Chandra, energy range not given.}
\tablenotetext{j}{\citet{hanson97}.}
\tablenotetext{k}{\citet{dunne03}, from ROSAT.}
\tablenotetext{l}{\citet{felli84}.}
\tablenotetext{m}{\citet{tapia03} and \citet{carraro04}.}
\tablenotetext{n}{\citet{seward79}, Einstein, 0.5 $-$ 3 keV.}
\tablenotetext{o}{\citet{okumura00}.}
\tablenotetext{p}{\citet{townsley04}, Chandra, 0.5 $-$ 8 keV.}
\tablenotetext{q}{\citet{smith78}.}
\tablenotetext{r}{\citet{piatti98} and \citet{cm04}.}
\tablenotetext{s}{\citet{melnick85} and \citet{brandl96}.}
\tablenotetext{t}{\citet{norci95}, from ROSAT, 0.1 $-$ 2.4 keV. }
\tablenotetext{u}{\citet{parker92}.}
\tablenotetext{v}{\citet{maclow98}, ROSAT, 0.5 $-$ 2.4 keV.}
\tablenotetext{w}{\citet{walborn92}.}
\tablenotetext{x}{\citet{oey95}.}
\tablenotetext{y}{\citet{dunne01}, ROSAT, 0.5 $-$ 2.4 keV.}
\tablenotetext{z}{\citet{oey96}, Chandra, 0.1 $-$ 2.4 keV.}
\tablenotetext{\rm aa}{\citet{oey97}.}
\tablenotetext{\rm bb}{\citet{bouret03}.}
\tablenotetext{\rm cc}{\citet{stevens03}.}
\tablenotetext{\rm dd}{\citet{relano02}.}
\tablenotetext{\rm ee}{\citet{tschoke00}, from ROSAT, 0.1 $-$ 2.4 keV.}
\tablenotetext{\rm ff}{\citet{martin92}.}
\tablenotetext{\rm gg}{\citet{williams95}.}
\tablenotetext{\rm hh}{\citet{skillman88}.}
\tablenotetext{\rm ii}{\citet{kuntz03}, Kuntz (2004) private communication.}
\tablenotetext{\rm jj}{\citet{bosch02}.}
\end{deluxetable}